\documentclass[fleqn,10pt]{wlscirep}
\usepackage[utf8]{inputenc}
\usepackage[T1]{fontenc}

\newcommand \dd[1]  { \,\textrm d{#1} }   
\newcommand \Ep  { \varepsilon_{\phi} }

\newcommand \En  { \varepsilon_K }
\newcommand \Es  { \varepsilon_S }

\graphicspath{ {figs/} }

\DeclareMathOperator{\sech}{sech}

\title{Ultrafast Electron Holes in Plasma Phase Space Dynamics}

\author[1,2,*]{Seyyed Mehdi Hosseini Jenab}
\author[3]{Gert Brodin}
\author[4]{James Juno}
\author[5]{Ioannis Kourakis}

\affil[1]{Department of Physics, Chalmers University of Technology, G\"oteborg, SE-412 96 ,Sweden}
\affil[2]{Department of Electrical, Computer \& Biomedical Engineering, Ryerson University, Toronto, M5B 2K3 ,Canada}
\affil[3]{Department of Physics,  Ume\aa \ University, Ume\aa, Sweden}
\affil[4]{Department of Physics and Astronomy, University of Iowa, Iowa City IA 54224, USA}
\affil[5]{Department of Mathematics, Khalifa University of Science and Technology, Abu Dhabi, UAE}

\affil[*]{mehdi.jenab@chalmers.se, mehdi.jenab@ryerson.ca}

\begin{abstract}
	Electron holes (EH) are localized modes in plasma kinetic theory which appear as vortices in
	phase space. Earlier research on EH is based on the Schamel distribution function (df).
	A novel distribution function is proposed here, generalizing the original Schamel df in a recursive manner.
	Nonlinear solutions obtained by kinetic simulations are presented, 
	with velocities twice the electron thermal speed.
	Using 1D-1V kinetic simulations, their propagation characteristics are traced and 
	their stability is established by studying their long-time evolution 
	and their behavior through mutual collisions.
\end{abstract}

\begin{document}

\flushbottom
\maketitle

\thispagestyle{empty}

\section*{Introduction}

	Plasma phase-space dynamics is tacitly characterized by the occurrence of electron holes,
	a term describing a localized plasma region where electrons are trapped
	by the electric potential stemming from their own self-generated density variation,
	as a localized
	electron depletion region occurs in a self-consistent manner.
	An electron hole is thus manifested as a localized ``trapped'' electron population
	traveling alongside an electrostatic potential disturbance\cite{schamel1986electron,schamel2015particle}.
	Electron-holes present two main characteristics \cite{hutchinson2017electron}:
	a localized positive potential structure which traps electrons,
	and	a symmetry in the electric potential profile around the peak.
	In addition, electron holes are a type of
	Bernstein, Greene, and Kruskal (BGK) mode\cite{bernstein1957exact}.
	Electron holes have been observed and studied
	in laboratory experiments\cite{saeki1979formation},
	in space measurements\cite{ergun1998fast, franz1998polar, matsumoto1994electrostatic, kojima1997geotail, deng2006observations}
	and in kinetic simulations\cite{eliasson2006formation}.

	In order to construct electron holes
	in a self-consistent manner within a kinetic model,
	one may either start with an arbitrary potential profile
	and then proceed by deriving the distribution function (df) of an electron hole,
	or, inversely, start with a predefined df for the trapped electrons
	and thus derive the associated potential profile.
	The former (integral equation) method,
	due to Bernstein, Greene and Kruskal \cite{bernstein1957exact}
	leads to an infinity of solutions whose dynamical stability is not prescribed.
	The latter (differential equation) method,
	suggested by Schamel \cite{schamel_1,schamel_2,schamel_3,schamel_5},
	is based on a parametrized \emph{df} (henceforth referred to as ``the Schamel df'')
	allowing one to prescribe the shape of the trapped population
	(i.e. by assigning a value to parameter $\beta$ associated with the inverse temperature of the trapped population).
	\textbf{Recently, Schamel df is extended by adding new parameters and hence resulted in 
	variety of new solutions.
	Note, most of the solutions, i.e. $\phi(x)$ 
	are undisclosed\cite{schamel2020two}. 
	In the case of double layers, the Schamel df provides solutions 
	which are much faster than the thermal velocity \cite{schamel1983analytical}.
	In fact, as the authors in Ref. \cite{schamel1983analytical} have predicted,
	a strong double layer (DL) solution is obtained as a limiting variant of a solitary hole;
	see also \cite{schamel1986electron} for details.
	}
	The Schamel method combined with the pseudopotential approach\cite{Sagdeev}
	may provide initial conditions for a controlled numerical investigation of EH dynamics\cite{jenab2018sagdeev}.
	Recent studies \cite{jenab2018sagdeev,jenab2019headon}
	have shown that the Schamel-pseudopotential approach
	can produce nonlinear solutions with Mach numbers $1.0 < M < 10.0$.

	However, only solutions in the range $1.0 < M < 3.0$ are found to be stable
	for long times \cite{jenab2018sagdeev}	and
	to survive mutual collisions \cite{jenab2019headon}.
	In other words, structures are destabilized as the Mach number increases.
	This has been suggested in other kinetic simulations\cite{turikov1984electron}.
	For very high Mach number ($M > 10$),
	the Schamel-pseudopotential method
	can not provide any solutions even for
	a wide range of $\beta$ (values)\cite{jenab2018sagdeev,hutchinson2017electron}.
	The maximum speed for a soliton accompanied by an electron hole (SEH) is $M=1.307$ when
	using the pseudopotential appoach in the small-amplitude regime \cite{bujarbarua1981theory}.

	Despite these theoretical challenges, the existence
	of high-speed electron holes is a topic of intense
	study, first getting attention due to observations by
	the FAST satellite\cite{ergun1998fast,muschietti1999analysis,muschietti1999phase}.
	Saeki \textit{et al}\cite{saeki1979formation} studied electron holes
	experimentally using a Q-plasma machine and also via kinetic simulations;
	they reported structures moving at the electron thermal speed,
	which they identified as solitons.
	Solitons are nonlinear structures that can survive mutual collisions
	and are characterized by a phase shift during a collision
	\cite{nishida1980oblique, verheest2012head,
	nakamura1999oblique, marchant2002asymptotic, demiray2007interactions}.
	We note however, Saeki \textit{et al} did not consider the phase 
	shift separating the hole trajectories before and after collisions.
	It is interesting to point out that fast (large Mach number)
	localized structures have also been predicted recently,
	in the form of supersolitons (supernonlinear waves); see e.g. \cite{Dubinov,Verheest2013,Saha2020}.
	Nonetheless, it is important to realize that these structures are distinct 
	in both their structural characteristics (shape) and in the physical mechanism underlying their formation.
	(An interested reader is referred to the above references for details.)

	The aim of this study is to characterize high-speed electron holes
	by establishing their occurrence in a kinetic framework,
	and by investigating their stability profile
	and probing their soliton-like features.
	For this purpose, a novel distribution function (\emph{df}),
	the `\texttt{ELIN} df', is introduced as
	a generalization of the Schamel df.
	The \texttt{ELIN} df adjusts the distribution function
	of the trapped population of electrons by relying on a dynamically varying parameter $\beta$
	so that its moments can fit a predetermined curve
	and all of the desired featured of the Schamel df are retained,
	such as consistency and smoothness in both spatial and velocity spaces
	inside the trapped region.

	To show the stability of our nonlinear solutions,
	three series of simulations are reported.
	Firstly, by considering the long-time evolution of an initial condition
	we will confirm the stability of the solution's profile during propagation,
	thus establishing them as solitary waves.
	Then, two types of mutual collisions are reported, i.e.
	head-on collisions (with no overlapping in  velocity space)
	and  overtaking collisions  (moving in parallel and with overlapping).
	The aforementioned phase shift through collisions has also been investigated,
	to corroborate the fact that electron holes behave as  solitons.

\section*{Results}

	\subsection*{long-term evolution}

\begin{figure}
	\centering
		\includegraphics[width=0.75\textwidth]{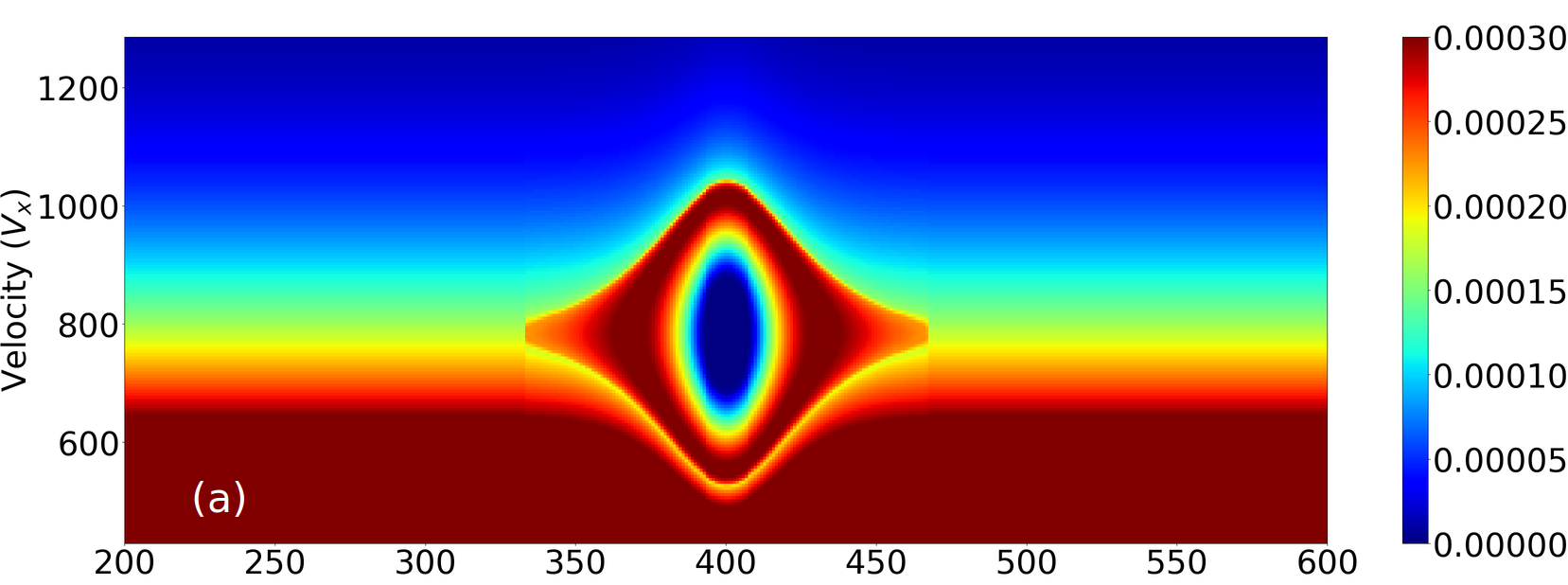}\\
		\includegraphics[width=0.75\textwidth]{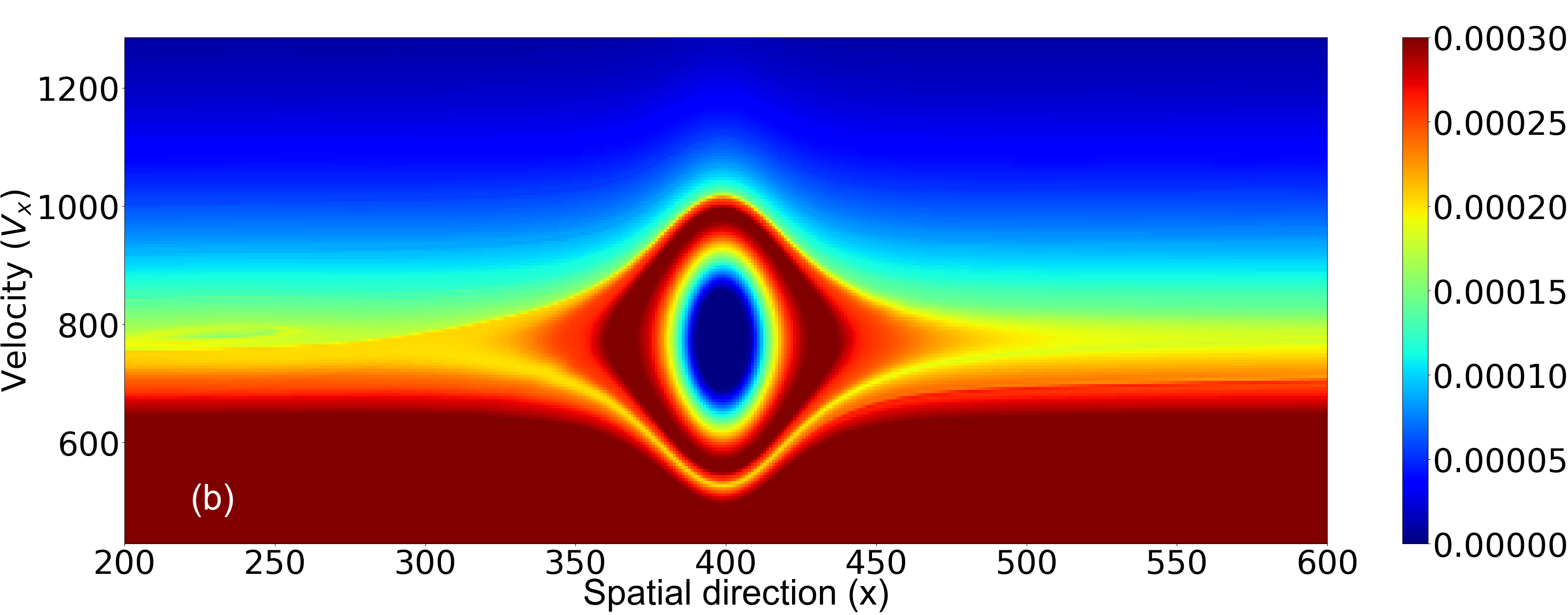}
		\caption{
		The electron phase space is shown for for   case $EH1$
		(a) at the initial step and (b) at $\tau = 12$.}
		\label{Fig_propagation_M45_EH}
	\end{figure}

\begin{figure}
	\centering
		\includegraphics[width=0.75\textwidth]{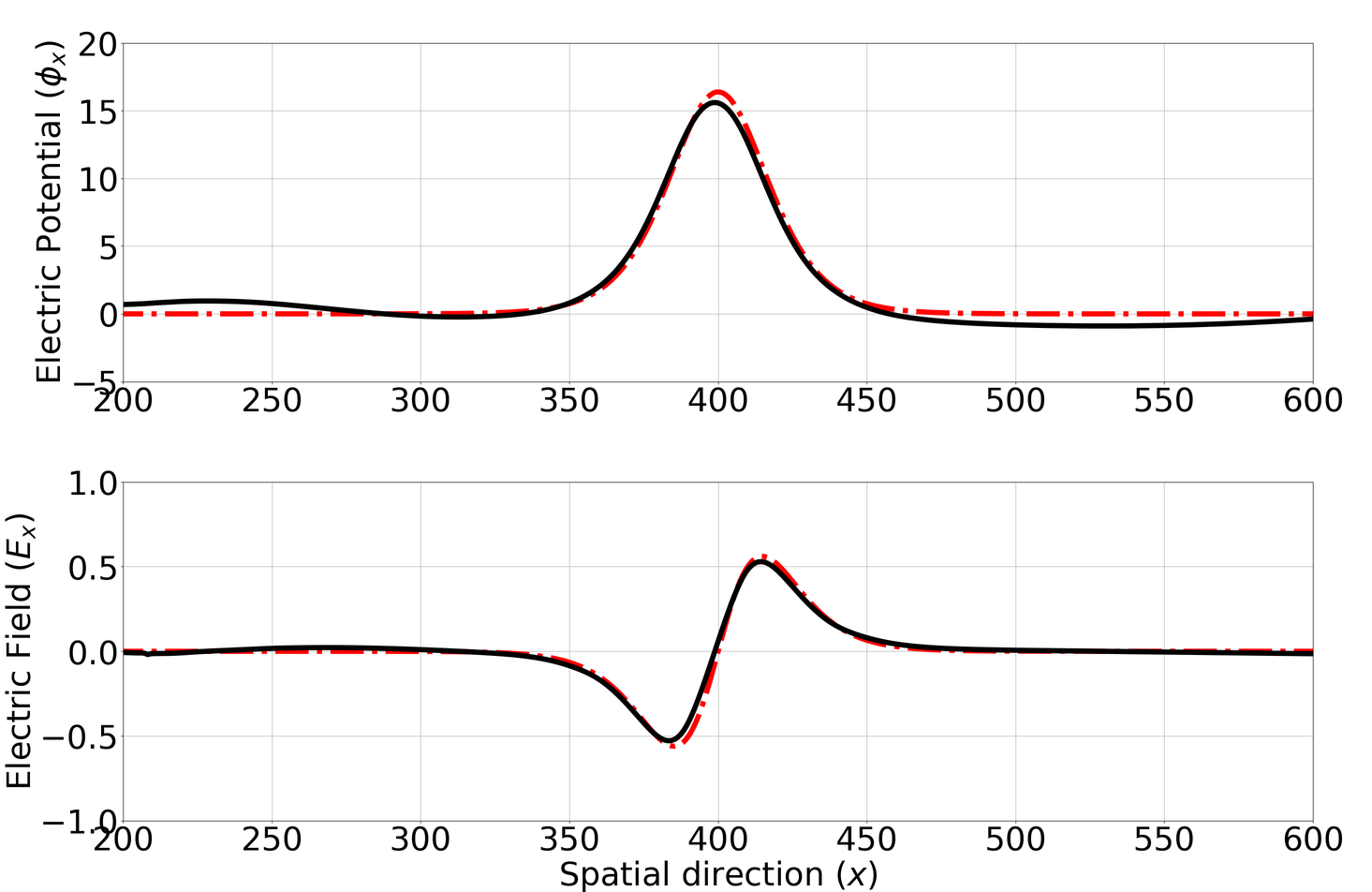}
		\caption{
		The electrostatic potential/E-field profile of $EH1$ is shown in the top/bottom panel.
		The initial condition
		i.e. at $\tau=0$ (red dotted curve) is compared with
		$\tau=12.0$  (solid black curve), showing a good agreement and
		hence confirming the stability of EHs during long-time propagation.
		}
		\label{Fig_propagation_M45_profile}
	\end{figure}

	Figs. \ref{Fig_propagation_M45_EH} and \ref{Fig_propagation_M45_profile}
	display the temporal evolution
	of $EH1$. 
	The initial condition and the last step of temporal evolution
	can be compared and show that the overall shape of the electron hole (Fig.\ref{Fig_propagation_M45_EH})
	and the corresponding potential or field profile
	(Fig.\ref{Fig_propagation_M45_profile})stay unperturbed.

	\subsection*{head-on collision}
\begin{figure}
	\centering
		\includegraphics[width=0.75\textwidth]{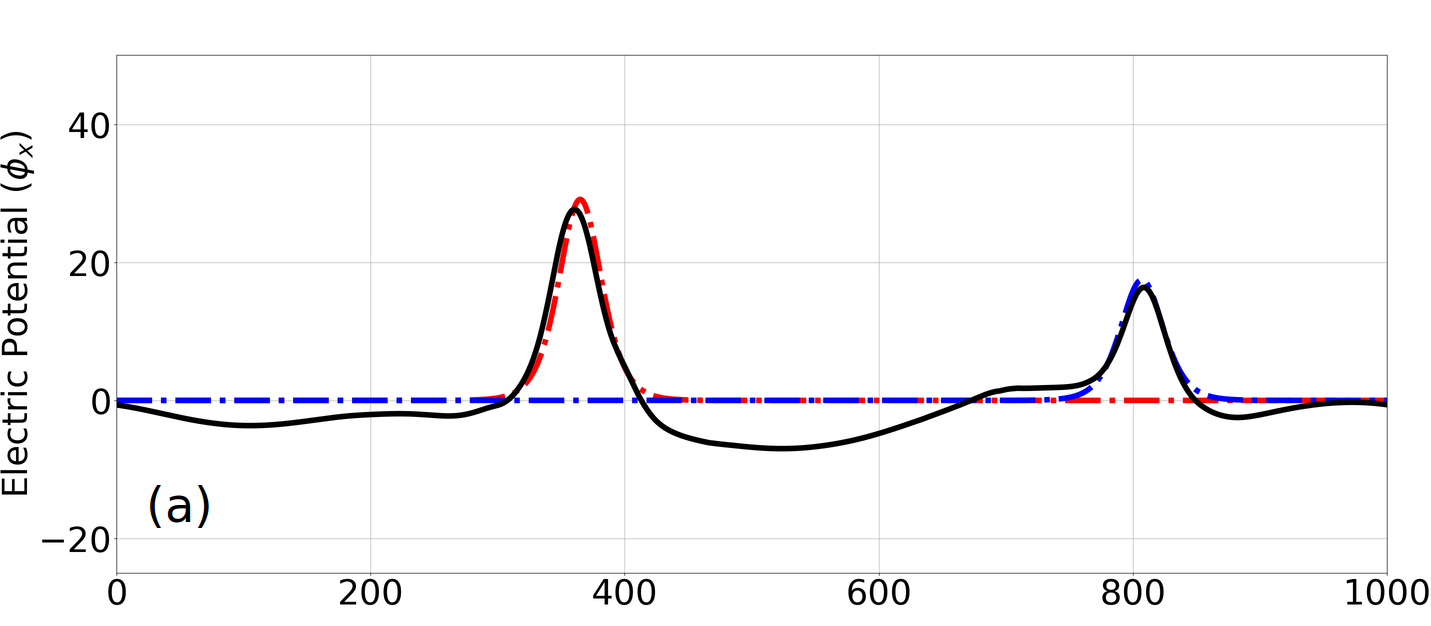}\\
		\includegraphics[width=0.75\textwidth]{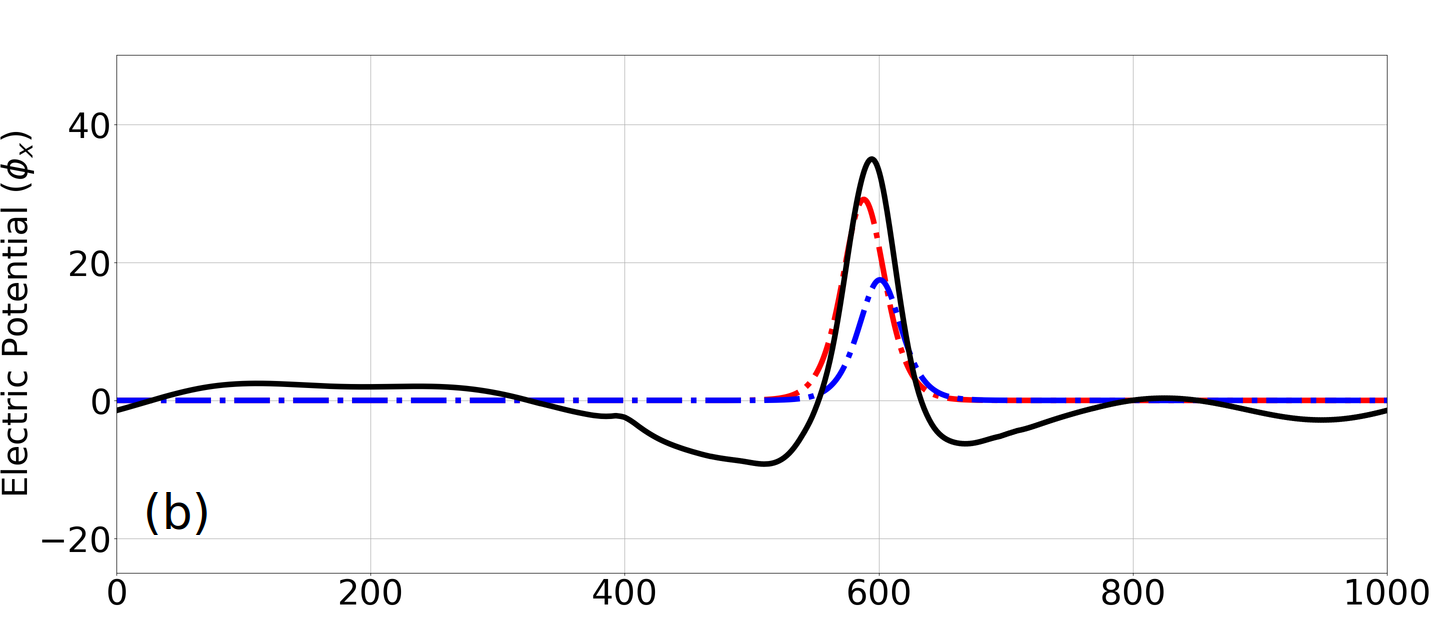}\\
		\includegraphics[width=0.75\textwidth]{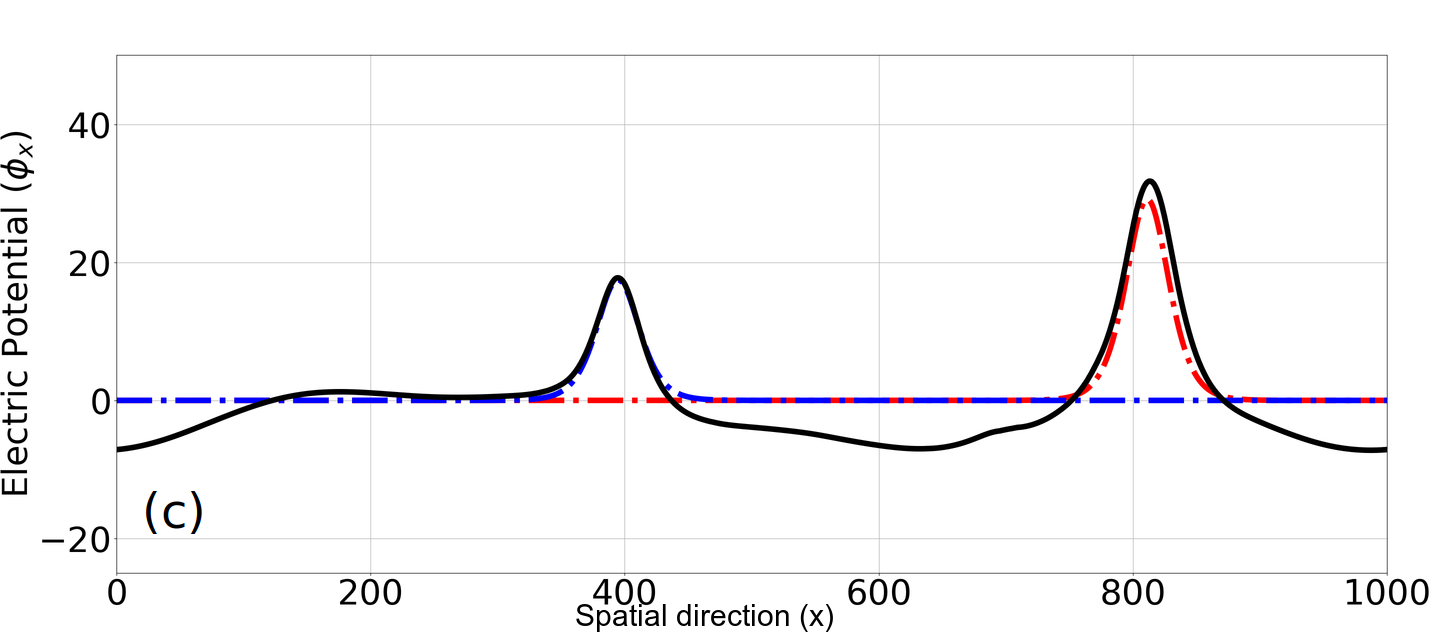}
		\caption{The electrostatic potential profile of $EH1$ and $EH2$ is shown
		at different snapshots around a head-on collision, namely
		a) before ($\tau=1.3$),
		b) during ($\tau=1.6$)
		and c) after ($\tau=1.9$) the collision.
		Dotted lines represent the initial condition for each of the solitary wave
		as if they are propagating without any numerical noise or collisions.
		Red/blue is for EH1/EH2 which is propagating to the right/left.
		After the collision,
		the overall shape and velocity of the solitary wave remains intact.
		}
		\label{Fig_headon_M45}
	\end{figure}

	Fig. \ref{Fig_headon_M45} depicts a head-on collision
	between $EH1$ and $EH2$. 
	After the  collision ($0 < \tau < 2$), both solutions
	keep their shape and velocity compared to their initial state.
	Note that due to their large velocity,
	they are well-apart in the velocity direction, i.e. there is no overlapping,
	and hence their collision on the phase space consists
	of two electron holes passing each other without much  interaction.
	Both electron holes follow their unperturbed trajectories after the collision,
	hence no phase shift is observed.

	\subsection*{overtaking collision}
\begin{figure}
	\centering
		\includegraphics[width=0.65\textwidth]{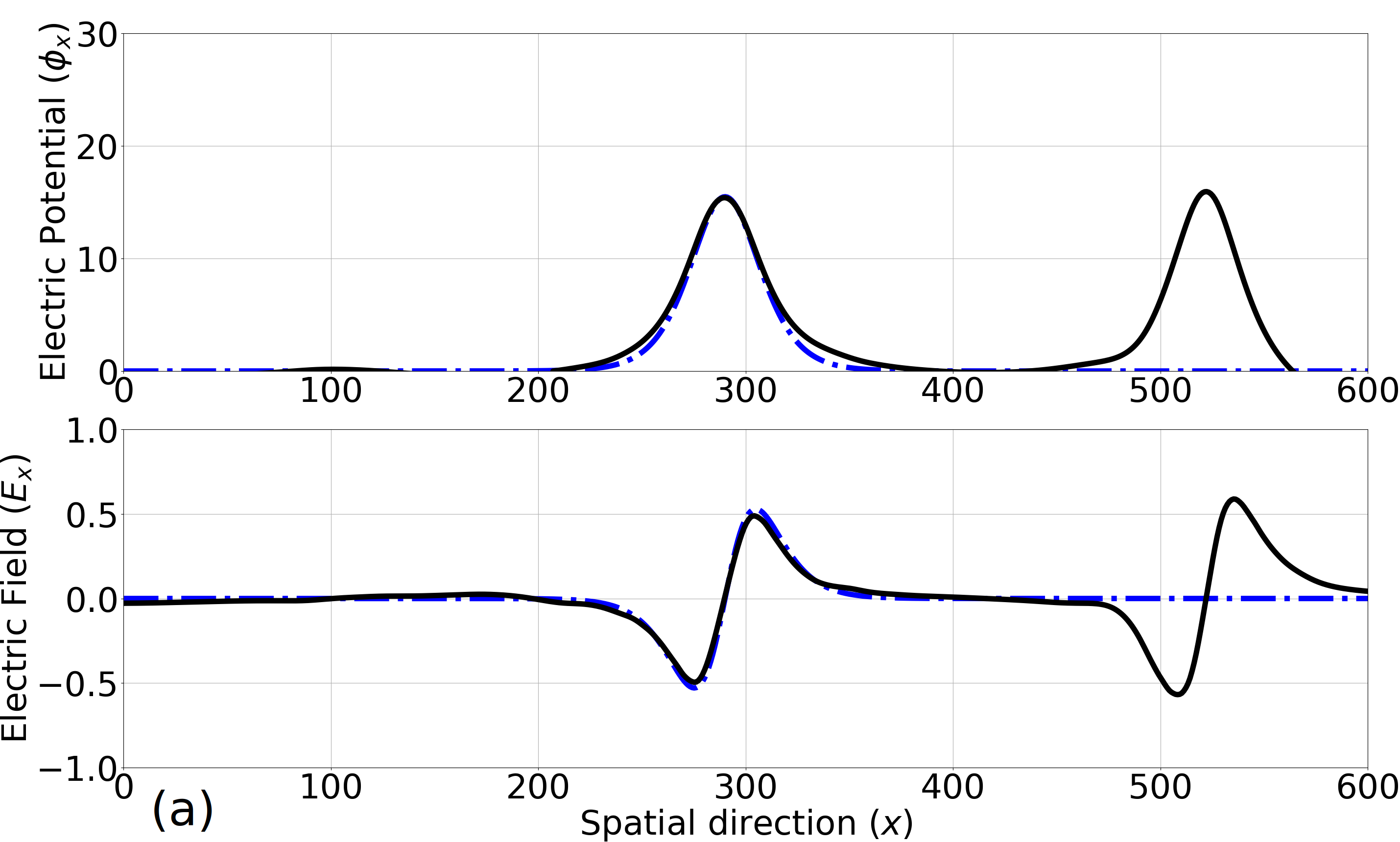}\\
		\includegraphics[width=0.65\textwidth]{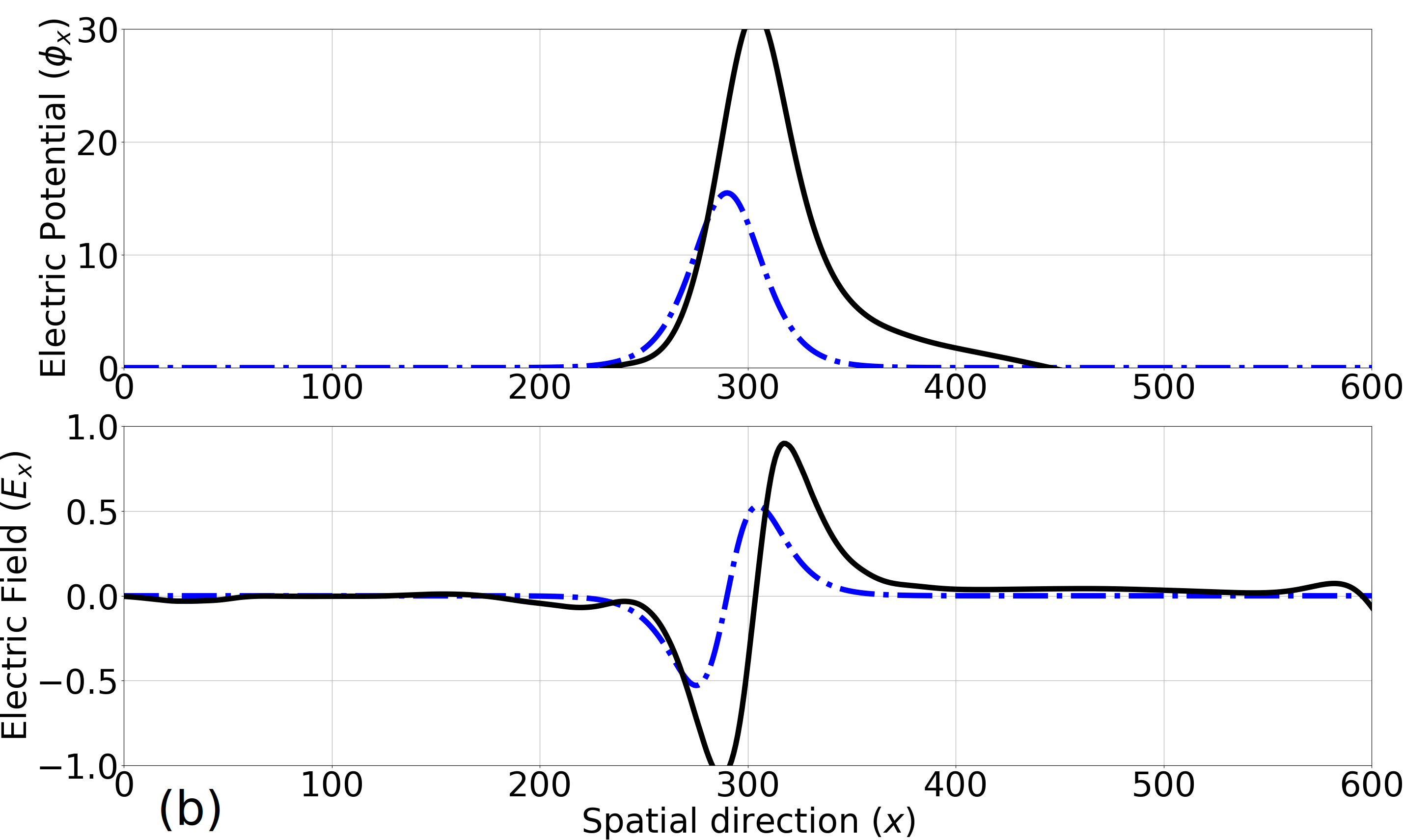}\\
		\includegraphics[width=0.65\textwidth]{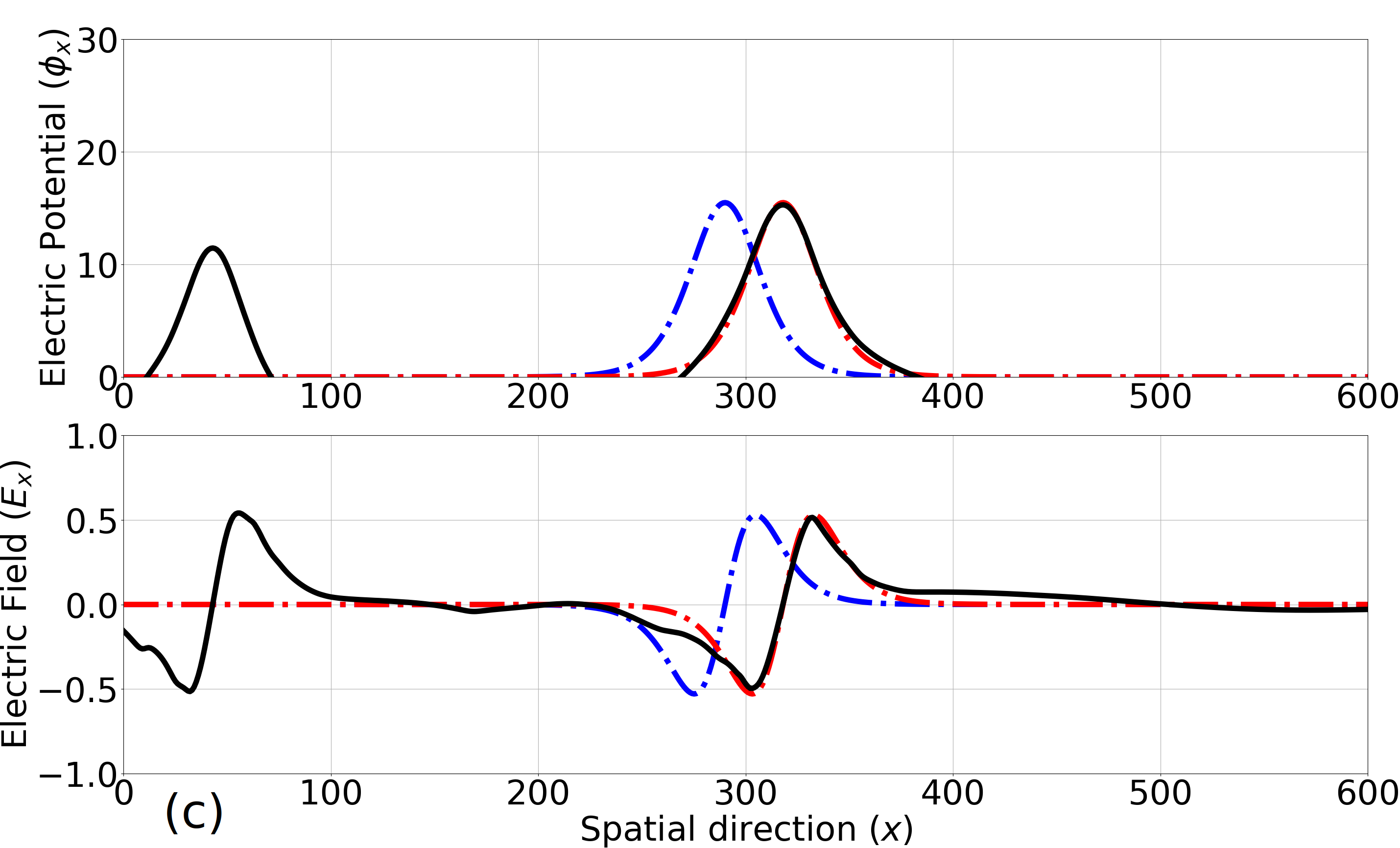}
		\caption{An overtaking collision between $EH1$ and $EH3$ is presented by plotting the 
		electrostatic potential and the electric field profile in the co-moving frame of $EH1$
		at three snapshots: 
		a) before ($\tau=2.35$),
		b) during ($\tau=3.20$) and
		c) after ($\tau=4.27$) the collision.
		The dotted curves show the fitted profile (${sech}^2$
		before (blue) and after (red) the collision, for $EH1$.
		A shift in the position of the first EH can be witnessed
		(note the difference between the red and the blue curves)
		manifesting a phase shift, as intuitively expected.}
		\label{Fig_overTaking_Ex}
	\end{figure}

	Although the previous simulations
	demonstrate the stability of these
	EHs, the strongest test of the stability is their interaction via an overtaking collision
	when they  overlap in the velocity direction.
	In an overtaking simulation, we have used two EHs e.g. $EH1$ and $EH3$.
	Fig. \ref{Fig_overTaking_Ex} presents the temporal evolution
	of electric field/potential
	around the collision time $\tau=3.2$ in a frame moving with $M=45$.
	Both EHs  survive the collision, and their respective velocity stays  the same.
	Focusing on $EH1$, displacement can be witnessed
	after the collision.
	A phase shift can be measured by comparing EH profile
	with the red line, which is an extrapolation
	of an unperturbed path of this EH.
	This displacement is similar to the well-known effect of ``phase shift''
	which observed to happen in mutual collisions of solitons
	\cite{nishida1980oblique, verheest2012head,
	nakamura1999oblique, marchant2002asymptotic, demiray2007interactions}.

\begin{figure}
	\centering
		\includegraphics[width=0.75\textwidth]{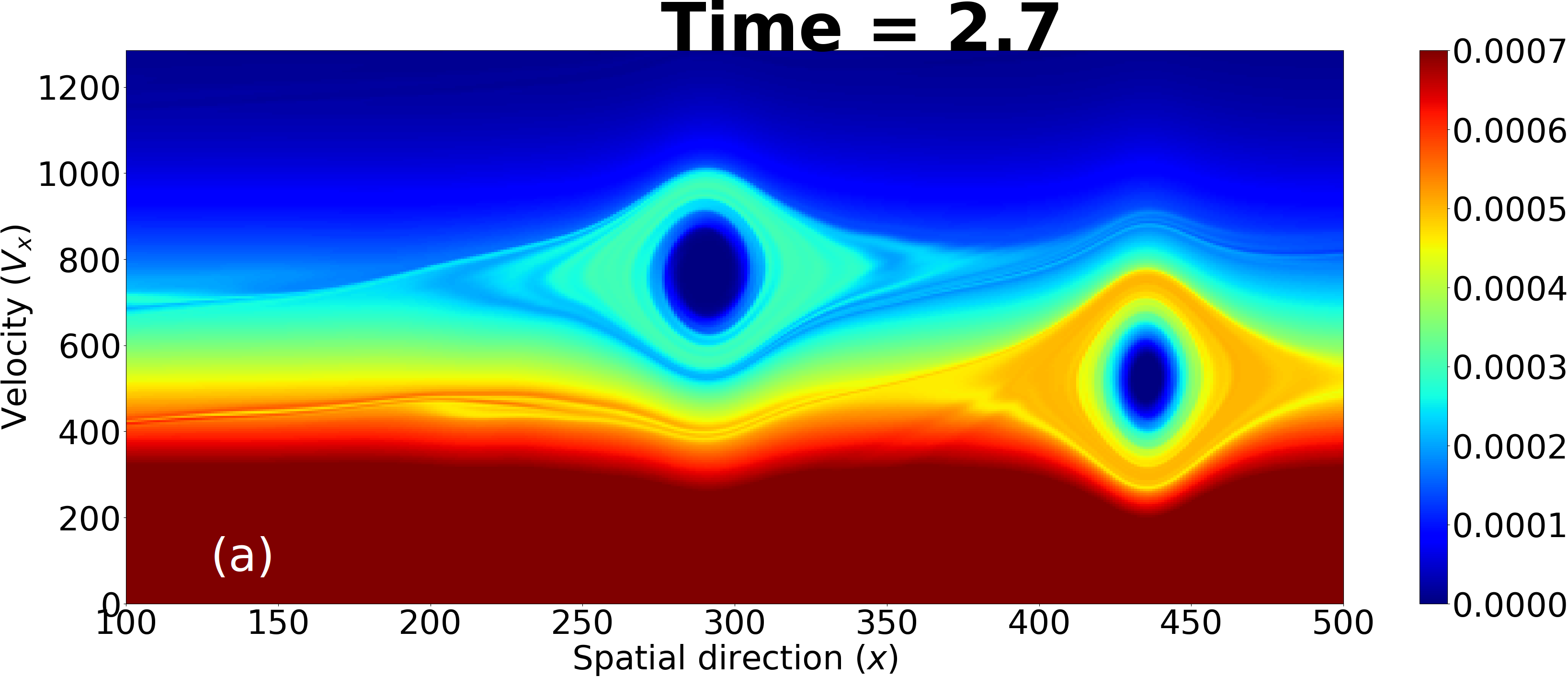}\\
		\includegraphics[width=0.75\textwidth]{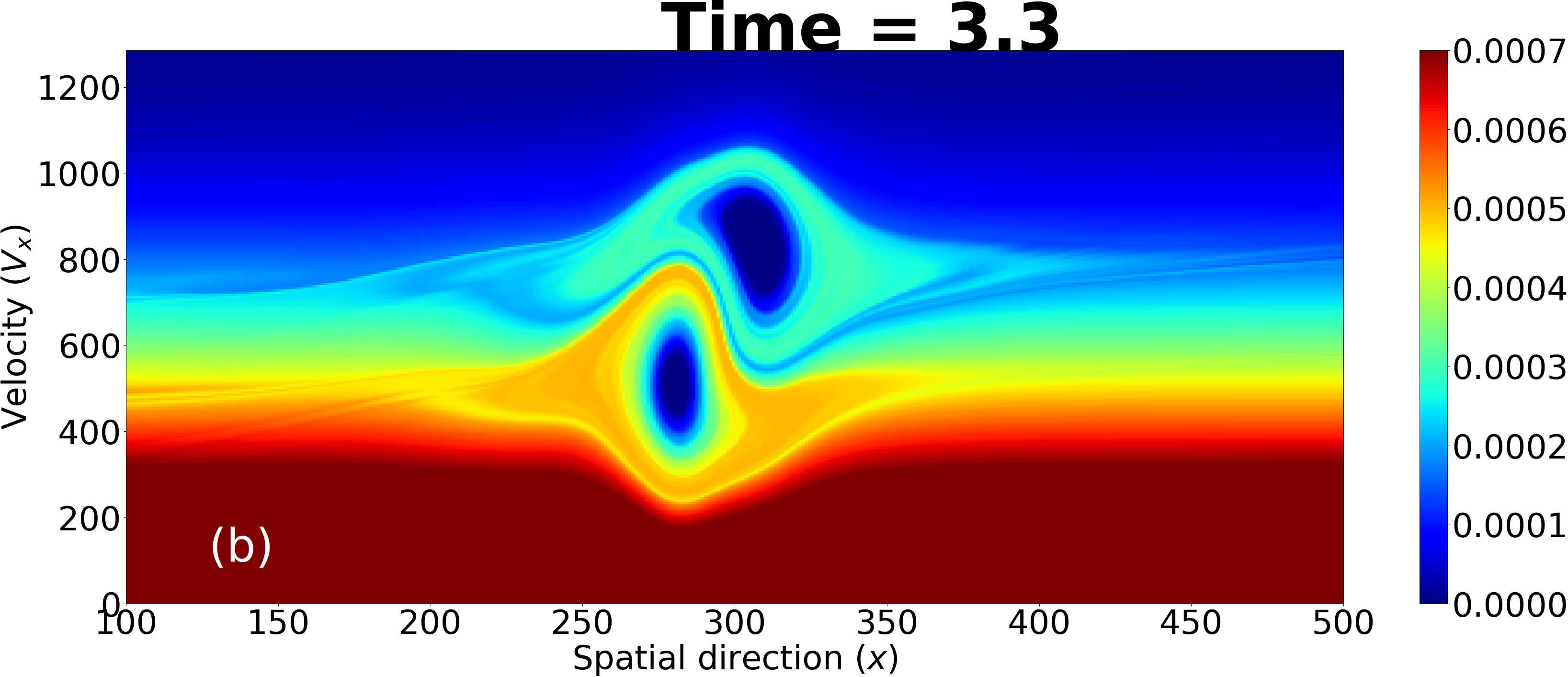}\\
		\includegraphics[width=0.75\textwidth]{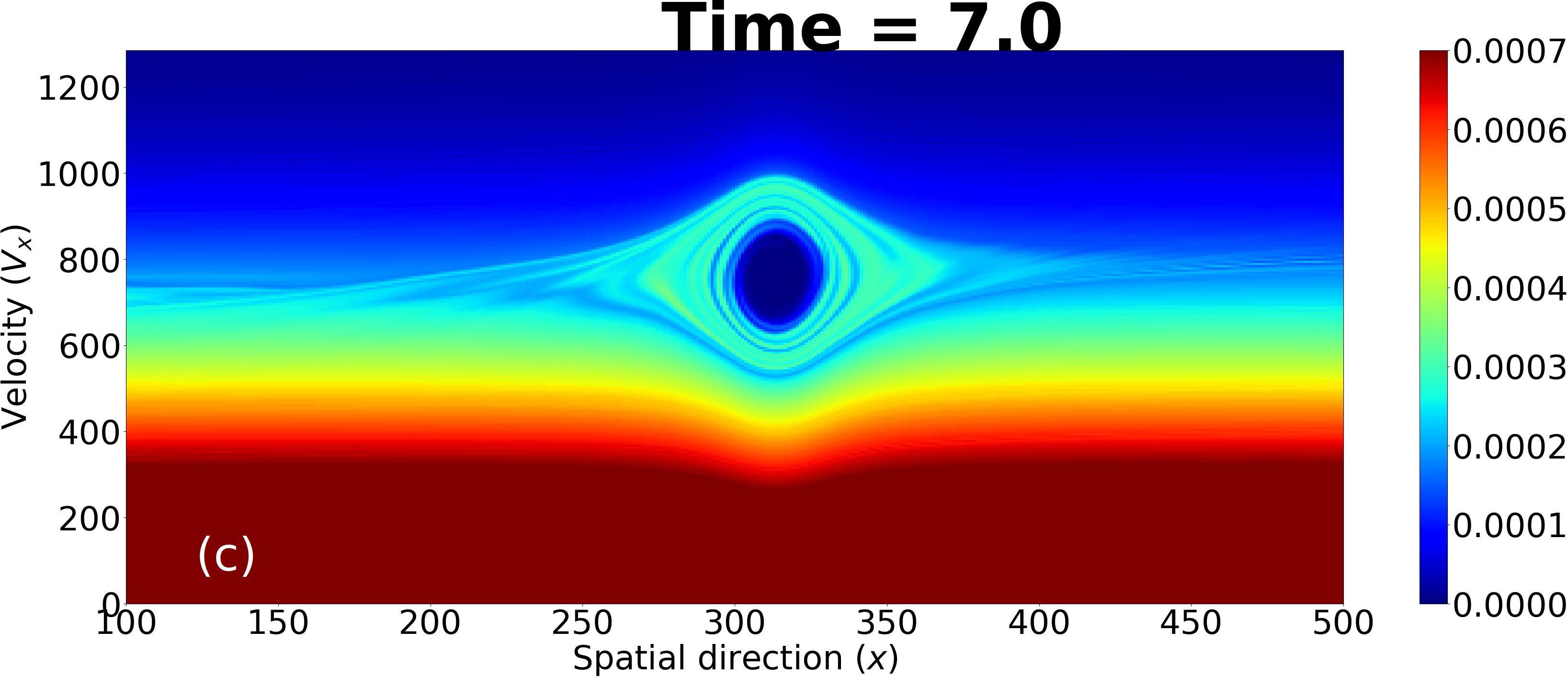}
		\caption{The electron phase space is presented for
		an overtaking collision between $EH1$ and $EH3$
		in the co-moving frame of $EH1$.
		There is a substantial overlap in velocity direction (a).
		During collision, the interaction is strong (b).
		After the collision, $EH1$ reappears un-altered in (c).
		}
		\label{Fig_overTaking_df}
	\end{figure}

	We show in Fig.\ref{Fig_overTaking_df} the electron df during the overtaking collision,
	which demonstrates the considerable interaction between the EHs during the collision
	and their overlapping on velocity direction.
	Yet after the collision the $EH1$ is largely unperturbed,
	modulo the observed phase shift.
	Interestingly, data fitting has shown that the $\sech^2$ curve form
	approximates the numerical data better than any other exponent,
	including the (expected, arguably) $\sech^4$ form (see Eq. 39 in \cite{schamel_3}).

\section*{Discussion}
	In summary,
	we have provided a method
	to produce high-speed nonlinear solutions which move at a speed 
	beyond the electron thermal speed.
	We showed that these electron holes are stable, retain their profile through 
	collisions and remain so in the entire duration of the simulation.
	For mutual collisions with considerable overlap in the velocity direction,
	the EHs display a ``phase shift''
	This phase shift represents a signature
	of soliton behavior and hence suggests that
	these EHs can be considered as solitons (at least approximately).
	This has been suggested for much lower-speed EHs before\cite{saeki1979formation}
	but without the observed ``phase shift'' reported here.


\section*{Methods}

	\begin{figure}[ht]
	\centering
		\includegraphics[width=0.75\textwidth]{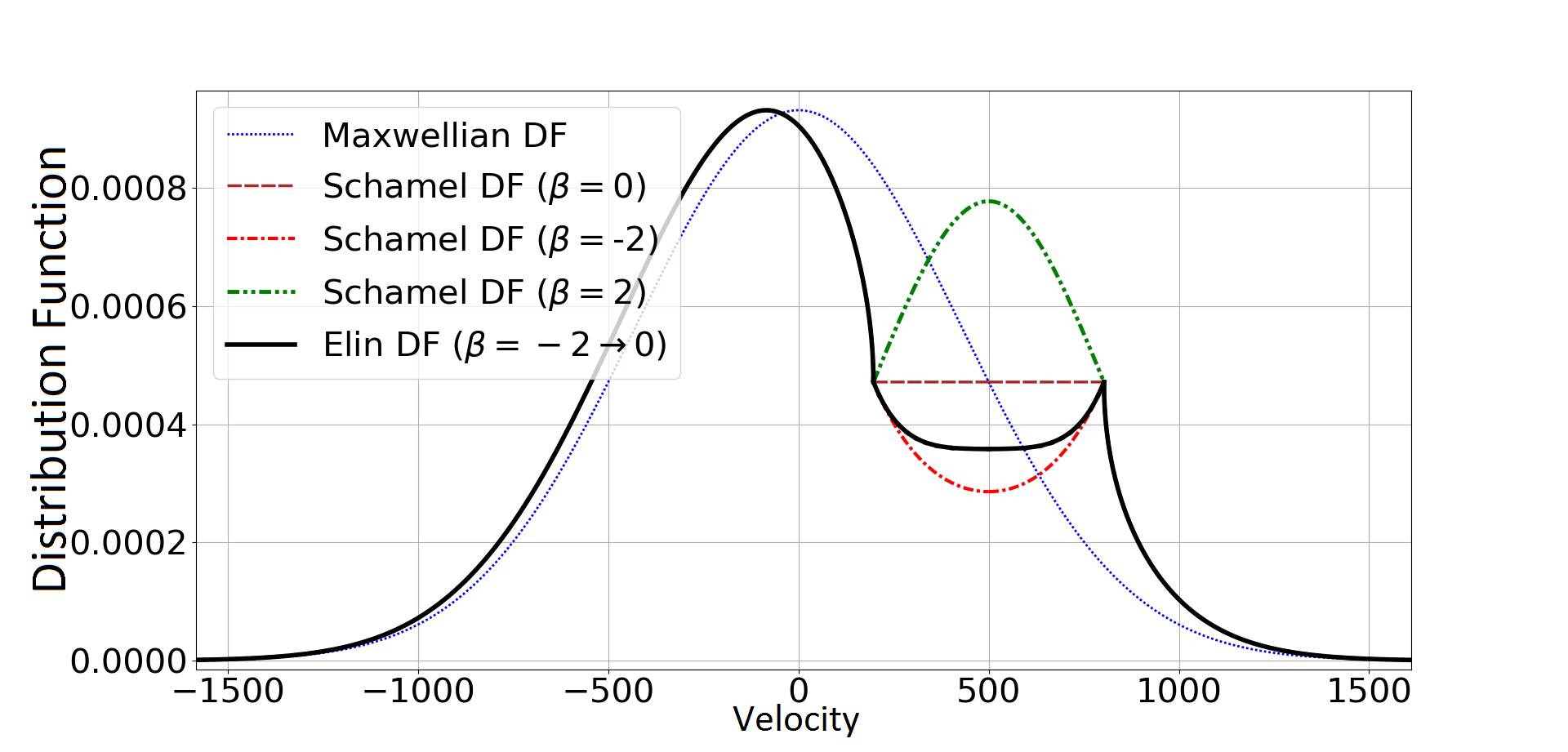}
		\caption{Different distribution functions for the trapped electron population are presented.
		The Maxwellian df (in the absence of trapped particles)
		is shown for sake of comparison (blue, thin dotted line).
		Three shapes of the Schamel df are displayed,
		namely flat (brown dashed, $\beta=0$), hollow (red, dashed-dotted, $\beta=-2$) and a
		bump (green, dashed-dotted, $\beta=2$),
		for $\phi = 25$.
		The \texttt{ELIN} df (black thick line) is shown
		when ten carving ($\phi_1 = 2.5$, $\phi_2=5$, $\phi_3=7.5$, ...,  $\phi_{10} = 25$)
		is carried out with their corresponding
		$\beta$ ($\beta_1 = -2, \beta_2=-1.8, \beta_3=-1.6, ..., \beta_{10}=0$). }
		\label{Fig_Elin}
	\end{figure}


	\subsection*{Equation set}
	The scaled Vlasov-Amp\`ere system of equations forming the basis of our simulation
	reads:
	\begingroup\makeatletter\def\f@size{9}\check@mathfonts
	\def\maketag@@@#1{\hbox{\m@th\large\normalfont#1}}%
	\begin{align}
		\frac{\partial f_s(x,v,t)}{\partial t}
		+ v \frac{\partial f_s(x,v,t)}{\partial x}
		&+  \Upsilon_s E(x,t) \frac{\partial f_s(x,v,t)}{\partial v} 
		= 0, \\
		\frac{\partial E(x,t)}{\partial t} & =\sum q_s J_s(x,t)
		\label{Vlasov}
	\end{align}\endgroup
	where $s = i,e$ represents the corresponding species,
	i.e. ions and electrons respectively.
	The factor $\Upsilon_s$ takes the values $\Upsilon_e= -1836$ and  $\Upsilon_i= 1$.
	The normalized charges are $q_e = -1$ and $q_i = 1$.
	The above equations are coupled by integrations for each species, viz.
	$J_s(x,t) = \int f_s(x,v,t) v dv$
	in order to form a closed set of equations for $J$, denoting the current (contribution) generated by
	by species $s$. To derive the above (dimensionless) equations,
	all physical quantities were normalized to suitable scales related with ionic parameters, i.e.
	mass ($m_s$) was divided by the ion mass ($m_i$),
	temperature ($T_s$) by ion temperature ($T_i$),
	charge ($q_s$) by the elementary charge ($e$),
	time ($\tau$) by the ion plasma period
		($\omega_{pi}^{1/2}  = {\big(\frac{n_{i0} e^2}{m_i \epsilon_0}\big)^{-\frac{1}{2}} }$),
	and 	length  ($L$)  by the ion Debye length
		($\lambda_{Di} = \sqrt{ \frac{\epsilon_0 K_B T_i}{n_{i0} e^2}   }$).
	Here, $K_B$ is Boltzmann's constant and
	$\epsilon_0$ is the permittivity of free space.

	\subsection*{Simulation code}
	We have employed the \texttt{Gkeyll} simulation framework \cite{hakim2020continuum}
	to solve the Vlasov-Ampere system of equations \cite{juno2018discontinuous, hakim2019conservative, hakim2020alias}.
	\texttt{Gkeyll} discretizes
	the equations
	using the discontinuous Galerkin finite element method in space,
	with
	a strong stability-preserving
	Runge-Kutta method
	in time.
	We have adopted a piecewise cubic Serendipity Element space for the basis expansion\cite{arnold2011serendipity}
	(further details can be found in Refs. \cite{juno2018discontinuous} and \cite{hakim2020alias}).
	The \texttt{Gkeyll} method has been compared to the standard PIC method, where it was demonstrated that the effective phase space resolution of the method is very high, permitting detailed studies of df dynamics. Such high accuracy is of paramount importance for the resolution of EH dynamics in phase space\cite{juno2020vlasovpic}.

	\subsection*{Parameters}
	In our study, the temperature and mass ratio are
	$\frac{T_e}{T_i}=100$  and  $\frac{m_i}{m_e}=1836$.
	The initial distribution function $f_0$
	is considered to be the Maxwellian df ($=D_m$).
	The size (length) of the simulation box is $l=1000$ in the $x$-direction.
	In the $v$ direction for each species, we have different limits:
	for the electrons we have $v=(-6, 6) v_{th_e} = (-2571,2571)$
	and for the ions we have $v=(-10, 10)$, where
	$v_{th_e} = \sqrt{\frac{T_e}{T_i}\frac{m_i}{m_e}}\approx428.5$
	is the electron thermal velocity.
	The number of grid cells in   each direction is $n_X = 2000$,
	$n_V = 1000$ for both  electrons and ions.
	The time step $\dd \tau \approx  10^{-5}$ is chosen in order to
	fulfill Courant-Friedrichs-Lewy (CFL) condition\cite{courant1928partiellen,courant1967partial}.

	The electron hole speed ($v_{EH}$) is expressed
	by the ``Mach number'', which is defined as the ratio $M = \frac{v_{EH}}{c_s}$, where
	$c_s = \sqrt{1+\frac{\gamma_e T_e+\gamma_i T_i}{m_i}}$ is the ion sound speed.
	Assuming $\gamma_e = \gamma_i = 3$ (heat capacity ratio), $T_e = 100 T_i$ and $m_i = 1$,
	the ion sound speed in our simulations is $c_s =\sqrt{304} \approx 17.43$.

	\subsection*{Iterative method to find stable solutions}
	Our method follows the BGK method and starts
	by adopting  an arbitrary function
	for the electrostatic potential $\big(\phi(x)\big)$
	and by choosing the value of the electron hole speed $(v_{EH})$.
	We then use the \texttt{ELIN} df to produce
	the electron distribution function. Given that the potential
	profile provides the charge density $(\rho(x))$, and using
	the Schamel df for the ions to obtain $n_i(x)$,
	we then use the total charge density (profile)
	$n_e(x) = \rho(x)-n_i(x)$
	as a “guiding equation” for the \texttt{ELIN} df
	and thus  construct the electron hole.
	We have adopted, to start with, the simplest form of potential profile suggested for electron holes
	i.e. $\phi = A \sech^p (x/L)$ in which $p=2$
	and $A$ and $L$ are the EH amplitude and length, respectively.
	The amplitude and length (values) are chosen randomly;
	however the system will damp/break	the forced profile if it is not close-enough to a self-
	consistent nonlinear solution. The resulting electron
	hole may have different size and velocity, but with
	an iterative process, one can find the combination of
	$\{A, L\}$ for which the solution will be stable enough for a specific (chosen) velocity value.
	Since we are not aware of the nonlinear dispersion relation,
	i.e. a relationship between $\{A, L, M, p\}$ for the exact
	nonlinear solution(s),
	a sequence of trials
	is performed to iterate to the correct combination of
	$\{A, L, p\}$ for a given $M$.
	In the simulations presented here three electron holes were studied, e.g.
	\begin{itemize}
		\item $EH1$: $M= 45, A=19,  L=22.5, p=2$
	    \item $EH2$: $M=-40, A=9.5, L=22.5, p=2$
	    \item $EH3$: $M= 30, A=19,  L=22.5, p=2$
	\end{itemize}

\subsection*{Elin DF method to construct electron holes}

	In order to explain our novel distribution function approach, firstly
	we need to represent the Schamel distribution function 
	in energy-dependent format. 
	Here we briefly discuss this, more details can be found in the reference\cite{jenab2018sagdeev}.
	Schamel approach devides the distribution function into two parts, namely
	free and trapped particles which are separated by a separatrix.

	Focusing on the free particles, 
	the following steps are taken to determine their distribution function ($f_f$), 
	assuming a pulse moving with a velocity ($v_{EH}$) in the laboratory frame:
	\begin{enumerate}
		\item the shifted kinetic energy is found in the co-moving frame:
		$\varepsilon'_{K_{sh}}=|\En'-\Ep|$
		where $\Ep=q\phi$, $\En' = \frac{1}{2} \frac{m}{T} v'^2$ and $v'= v - v_{EH}$ is the 
		velocity in the co-moving frame.
		
		\item the shifted kinetic energy is calculated in the laboratory frame:
		$\varepsilon_{K_{sh}} = \frac{1}{2} \frac{m}{T} v_{sh}^2$
		in which $v_{sh} = v'_{sh} + v_{EH}$ and subsequently  
		$v'_{sh}  = \textrm{sign} (v') 
		\sqrt{2\varepsilon'_{K_{sh}}/m}$
	\end{enumerate}
	Free particles fulfill the condition $\En'>\Ep$.
	Note that, in order to calculate the df at point $v$,
	we use the df at the point $v_{sh}$,
	which can be written as $f=D_g(\varepsilon_{K_{sh}})$ in energy format.
	Here, $v_{sh}$ presents the velocity of particles before their interaction 
	with the potential profile. 
	By	$D_g$ we denote a general distribution function satisfying the Vlasov equation, i.e. in principle any 
	function	depending on the constant(s) of motion.
	Here, the energy is used to construct a valid function.
	Well-known examples of $D_g$ are the Maxwell-Boltzmann df,  the $\kappa$ df 
	\cite{vasyliunas1968survey,pierrard2010kappa,summers1991modified,hellberg2009comment}
	and the Cairns\cite{cairns1995electrostatic} distribution function(s).

	In other words we trace the characteristics of the particle back in 
	phase space. 
	Then we use the value of df at $v_{sh}$ as the value of df for $v$ 
	since the df stays constant on the characteristics of Vlasov equation\cite{kazeminezhad2003vlasov}.

	The distribution function of trapped particles ($f_t$) 
	which are subject to the trapping condition ($\En'<\Ep$)
	can be achieved by following the steps below:
	\begin{enumerate}
		\item the shifted kinetic energy is found in the co-moving frame:
		$\varepsilon'_{K_{sh}}=|\En'-\Ep|$, 
		using a Maxwellian df on top of this kinetic energy with a coefficient $\beta$,
		will provide the shape of trapped distribution function:
		$f_{shape} =D_m(\beta \varepsilon'_{K_{sh}})= \exp (-\beta\varepsilon'_{K_{sh}})$

		\item In order to have continuity between trapped and free df 
		where they meet in the velocity direction, $f_{shape}$ is multiplied by  $f_{base} = D_g(\Es)$. 
		Hence $f_t = f_{base} \times f_{shape}$.
	\end{enumerate}
	Here, $ D_g(\Es)$ stands for the distribution function
	at the separatrix where $\En'=\Ep$ and 
	works as a constant value which can increase or decrese the $f_{t}$,
	in order to adjust it with the free distribution function. 
	The second component, $f_{shape}$ is velocity-dependent and is controlled by $\beta$.
	It may appear in three qualitative shapes, i.e. \emph{flat}, 
	a \emph{bump} or a \emph{hollow} curve, if $\beta = 0$, $\beta > 0$ or
	$\beta < 0$, respectively (see Fig. \ref{Fig_Elin}).

	Hence, the total form of the Schamel distribution function \cite{schamel_1} can be written
	in terms of the energy as: $f=a f(\En)$ in which $a$ is a normalization
	constant and 
		\begingroup\makeatletter\def\f@size{8.3}\check@mathfonts
		\def\maketag@@@#1{\hbox{\m@th\large\normalfont#1}}%
		\begin{equation}
			f(\En) =
				\left\{\begin{array}{lr}
					f_f = D_g(\varepsilon_{K_{sh}})
					&\textrm{if} \
					\En'>\Ep \\
					D_g(\Es)
					&\textrm{if} \
					\En'=\Ep \\
					f_t = D_g(\Es) D_m\Big(\beta |\En'-\Ep| \Big)
					&\textrm{if} \
					\En'<\Ep \\
				\end{array}\right.
			\label{Schamel_df}
		\end{equation}\endgroup

	One can understand the Schamel df as carving up a given general distribution function ($D_g$)
	around a particular velocity (hole velocity) and inserting a Maxwellian df
	with arbitrary temperature inside the hole ($f_{t}$).

	In the above representation of the Schamel df we used the analytical form of $\phi(x)$.
	However, one can equally use the discretizied form of $\phi(x)$ 
	(by deviding it into a $n$ intervals of $\Delta \phi$).
	Schamel df can then be retrieved by $n\rightarrow \infty$ ( $\Delta \phi \rightarrow 0$).
	In terms of simulation approach, these two methods are equal since even when using 
	the analytical approach, one had to use discretization for $\phi(x)$ and there is limit 
	on how small $\Delta \phi$ can get. 
	
	In other words, to generate distribution function ($f_f$ and $f_t$) for each 
	interval,  we only need update the value of  $f_{base}$ in our approach and
	repeat the process. This results in multiple carvings, each based on the 
	previous distribution function and it recursively progresses. 
	
	We assume $\phi = A \sech^2 (x/L)$ as  crude approach 
	(stablished by trial and error in the beginning), 
	and then we dicretize 
	the first half of $\phi(x)$ into $n$ intervals 
	in the following form ($\Delta \phi = \frac{A}{n}$):
	\begin{align*}
	 \phi (x_1) &= \Delta \phi \\
	 \phi (x_i) &= \phi (x_{i-1}) + \Delta \phi \\
	 \phi (x_n = \frac{L}{2}) &= A.
	\end{align*}
	The second half will be the same as the first half except for a simple inversion. 
	Hence we just build the first half of df and
	the second half is just simple inverted copy of it.
	
	In this approach $\beta$ can be changed for each interval, and this add 
	a new degree of freedom to the Schamel df. 
	We call this \texttt{ELIN} (rEcursiveLy extendable distribution for a trapped populatIoN)
	distribution function.
	The distribution function for each interval can be presented by the following equation.
	In which the $D_g$ (in Schamel df) is replaced by the distribution function $f_{i-1}$ 
	of previous interval and each interval has its own $\beta_i$:
	\begingroup\makeatletter\def\f@size{8.3}\check@mathfonts
	\def\maketag@@@#1{\hbox{\m@th\large\normalfont#1}}%
	\begin{equation}
		f( \phi_i) =
			\left\{\begin{array}{lr}
				f_f = f_{i-1}(\varepsilon_{K_{sh}})
				&\textrm{if} \
				\En'> q \Delta \phi_i \\
				f_{i-1}(\Es)
				&\textrm{if} \
				\En'= q \Delta \phi_i \\
				f_t = f_{i-1}(\Es) D_m(\beta_i(|\En'-\Ep|))
				&\textrm{if} \
				\En'< q \Delta \phi_i \\
			\end{array}\right.
	\label{Elin_df}
	\end{equation}\endgroup
	in which $f_0$ is the initial unperturbed df (here assuming Maxwellian df, i.e. $f_0 = D_m$).
	$\beta_i$ can change arbitrarily in order for moments of df to fit 
	a ``guiding equation'' (here, the equation for the electron density).
	To obtain a smooth distribution function in the $x$ direction,
	one can increase $n$ until the
	numerically-desired level of smoothness is achieved.
	An example of the \texttt{ELIN} df profile is presented at Fig. \ref{Fig_Elin}
	which shows 10 successive (carving) iterations 
	with $\beta$ approaching zero from below (negative side).
	Note that since $\beta$ originates from a continuous guiding equation, 
	hence their successive values follow a pattern and are not randomly chosen.

	To conclude, 
	we have introduced a new method for constructing electron holes within a kinetic framework, which
	relies on a successive multi-step extension of the Schamel df (here represented in energy-dependent form),
	i.e. the \texttt{ELIN} df method.
	The \texttt{ELIN} df adopts a continuously varying value for $\beta$, 
	in contrast to the Schamel df where  $\beta$ is a constant. 
	This extension provides an infinite number of parameters for the \texttt{ELIN} df, 
	which enables it to construct an electron hole for
	any given bell-shaped potential profile. 
	In our computational approach,
	the number of free parameters in the \texttt{ELIN} df is
	finite and equals the number of intervals ($n$). 
	We have adopted an iterative method (inspired by Newton’s iterative scheme),
	built on top of the \texttt{ELIN} df method, to find the stable solutions.
	Starting from an initial guess, in each iteration
	of this method firstly we use the \texttt{ELIN} df to build an electron hole
	and then utilize the Vlasov-Poisson simulation method
	to follow the temporal evolution of the electron hole for a short time. 
	We use the potential profile at the end of each
	iteration as an input for the next round of iteration. 
	After a few iterations, the initial and final potential profiles are close
	enough for this to be considered as a stable configuration, for closure. 
	Then, one can move on to longer-time numerical experiments,
	to investigate the long-time evolution of these localized structures and their behavior through mutual collisions.
	As a representative set,
	three stable solutions (i.e. EH1, EH2 and EH3; see above) have been reported in detail.


\begin{thebibliography}{10}
\urlstyle{rm}
\expandafter\ifx\csname url\endcsname\relax
  \def\url#1{\texttt{#1}}\fi
\expandafter\ifx\csname urlprefix\endcsname\relax\def\urlprefix{URL }\fi
\expandafter\ifx\csname doiprefix\endcsname\relax\def\doiprefix{DOI: }\fi
\providecommand{\bibinfo}[2]{#2}
\providecommand{\eprint}[2][]{\url{#2}}

\bibitem{schamel1986electron}
\bibinfo{author}{Schamel, H.}
\newblock \bibinfo{journal}{\bibinfo{title}{Electron holes, ion holes and
  double layers: Electrostatic phase space structures in theory and
  experiment}}.
\newblock {\emph{\JournalTitle{Phys. Reports}}} \textbf{\bibinfo{volume}{140}},
  \bibinfo{pages}{161--191} (\bibinfo{year}{1986}).

\bibitem{schamel2015particle}
\bibinfo{author}{Schamel, H.}
\newblock \bibinfo{journal}{\bibinfo{title}{Particle trapping: A key requisite
  of structure formation and stability of vlasov--poisson plasmas}}.
\newblock {\emph{\JournalTitle{Physics of Plasmas}}}
  \textbf{\bibinfo{volume}{22}}, \bibinfo{pages}{042301}
  (\bibinfo{year}{2015}).

\bibitem{hutchinson2017electron}
\bibinfo{author}{Hutchinson, I.~H.}
\newblock \bibinfo{journal}{\bibinfo{title}{Electron holes in phase space: What
  they are and why they matter}}.
\newblock {\emph{\JournalTitle{Physics of Plasmas}}}
  \textbf{\bibinfo{volume}{24}}, \bibinfo{pages}{055601}
  (\bibinfo{year}{2017}).

\bibitem{bernstein1957exact}
\bibinfo{author}{Bernstein, I.~B.}, \bibinfo{author}{Greene, J.~M.} \&
  \bibinfo{author}{Kruskal, M.~D.}
\newblock \bibinfo{journal}{\bibinfo{title}{Exact nonlinear plasma
  oscillations}}.
\newblock {\emph{\JournalTitle{Physical Review}}}
  \textbf{\bibinfo{volume}{108}}, \bibinfo{pages}{546} (\bibinfo{year}{1957}).

\bibitem{saeki1979formation}
\bibinfo{author}{Saeki, K.}, \bibinfo{author}{Michelsen, P.},
  \bibinfo{author}{P{\'e}cseli, H.} \& \bibinfo{author}{Rasmussen, J.~J.}
\newblock \bibinfo{journal}{\bibinfo{title}{Formation and coalescence of
  electron solitary holes}}.
\newblock {\emph{\JournalTitle{Physical Review Letters}}}
  \textbf{\bibinfo{volume}{42}}, \bibinfo{pages}{501} (\bibinfo{year}{1979}).

\bibitem{ergun1998fast}
\bibinfo{author}{Ergun, R.} \emph{et~al.}
\newblock \bibinfo{journal}{\bibinfo{title}{Fast satellite observations of
  large-amplitude solitary structures}}.
\newblock {\emph{\JournalTitle{Geophysical Research Letters}}}
  \textbf{\bibinfo{volume}{25}}, \bibinfo{pages}{2041--2044}
  (\bibinfo{year}{1998}).

\bibitem{franz1998polar}
\bibinfo{author}{Franz, J.~R.}, \bibinfo{author}{Kintner, P.~M.} \&
  \bibinfo{author}{Pickett, J.~S.}
\newblock \bibinfo{journal}{\bibinfo{title}{Polar observations of coherent
  electric field structures}}.
\newblock {\emph{\JournalTitle{Geophysical research letters}}}
  \textbf{\bibinfo{volume}{25}}, \bibinfo{pages}{1277--1280}
  (\bibinfo{year}{1998}).

\bibitem{matsumoto1994electrostatic}
\bibinfo{author}{Matsumoto, H.} \emph{et~al.}
\newblock \bibinfo{journal}{\bibinfo{title}{Electrostatic solitary waves (esw)
  in the magnetotail: Ben wave forms observed by geotail}}.
\newblock {\emph{\JournalTitle{Geophysical Research Letters}}}
  \textbf{\bibinfo{volume}{21}}, \bibinfo{pages}{2915--2918}
  (\bibinfo{year}{1994}).

\bibitem{kojima1997geotail}
\bibinfo{author}{Kojima, H.} \emph{et~al.}
\newblock \bibinfo{journal}{\bibinfo{title}{Geotail waveform observations of
  broadband/narrowband electrostatic noise in the distant tail}}.
\newblock {\emph{\JournalTitle{Journal of Geophysical Research: Space
  Physics}}} \textbf{\bibinfo{volume}{102}}, \bibinfo{pages}{14439--14455}
  (\bibinfo{year}{1997}).

\bibitem{deng2006observations}
\bibinfo{author}{Deng, X.} \emph{et~al.}
\newblock \bibinfo{journal}{\bibinfo{title}{Observations of electrostatic
  solitary waves associated with reconnection by geotail and cluster}}.
\newblock {\emph{\JournalTitle{Advances in Space Research}}}
  \textbf{\bibinfo{volume}{37}}, \bibinfo{pages}{1373--1381}
  (\bibinfo{year}{2006}).

\bibitem{eliasson2006formation}
\bibinfo{author}{Eliasson, B.} \& \bibinfo{author}{Shukla, P.~K.}
\newblock \bibinfo{journal}{\bibinfo{title}{Formation and dynamics of coherent
  structures involving phase-space vortices in plasmas}}.
\newblock {\emph{\JournalTitle{Phys. Reports}}} \textbf{\bibinfo{volume}{422}},
  \bibinfo{pages}{225--290} (\bibinfo{year}{2006}).

\bibitem{schamel_1}
\bibinfo{author}{Schamel, H.}
\newblock \bibinfo{journal}{\bibinfo{title}{Stationary solutions of the
  electrostatic vlasov equation}}.
\newblock {\emph{\JournalTitle{Plasma Physics}}} \textbf{\bibinfo{volume}{13}},
  \bibinfo{pages}{491} (\bibinfo{year}{1971}).

\bibitem{schamel_2}
\bibinfo{author}{Schamel, H.}
\newblock \bibinfo{journal}{\bibinfo{title}{Non-linear electrostatic plasma
  waves}}.
\newblock {\emph{\JournalTitle{Journal of Plasma Physics}}}
  \textbf{\bibinfo{volume}{7}}, \bibinfo{pages}{1--12} (\bibinfo{year}{1972}).

\bibitem{schamel_3}
\bibinfo{author}{Schamel, H.}
\newblock \bibinfo{journal}{\bibinfo{title}{Stationary solitary, snoidal and
  sinusoidal ion acoustic waves}}.
\newblock {\emph{\JournalTitle{Plasma Physics}}} \textbf{\bibinfo{volume}{14}},
  \bibinfo{pages}{905} (\bibinfo{year}{1972}).

\bibitem{schamel_5}
\bibinfo{author}{Schamel, H.}
\newblock \bibinfo{journal}{\bibinfo{title}{Role of trapped particles and waves
  in plasma solitons-theory and application}}.
\newblock {\emph{\JournalTitle{Physica Scripta}}}
  \textbf{\bibinfo{volume}{20}}, \bibinfo{pages}{306} (\bibinfo{year}{1979}).

\bibitem{schamel2020two}
\bibinfo{author}{Schamel, H.}
\newblock \bibinfo{journal}{\bibinfo{title}{Two-parametric, mathematically
  undisclosed solitary electron holes and their evolution equation}}.
\newblock {\emph{\JournalTitle{Plasma}}} \textbf{\bibinfo{volume}{3}},
  \bibinfo{pages}{166--179} (\bibinfo{year}{2020}).

\bibitem{schamel1983analytical}
\bibinfo{author}{Schamel, H.} \& \bibinfo{author}{Bujarbarua, S.}
\newblock \bibinfo{journal}{\bibinfo{title}{Analytical double layers}}.
\newblock {\emph{\JournalTitle{The Physics of Fluids}}}
  \textbf{\bibinfo{volume}{26}}, \bibinfo{pages}{190--193}
  (\bibinfo{year}{1983}).

\bibitem{Sagdeev}
\bibinfo{author}{Sagdeev, R.}
\newblock \bibinfo{journal}{\bibinfo{title}{Cooperative phenomena and shock
  waves in collisionless plasmas}}.
\newblock {\emph{\JournalTitle{Reviews of Plasma Physics}}}
  \textbf{\bibinfo{volume}{4}}, \bibinfo{pages}{23} (\bibinfo{year}{1966}).

\bibitem{jenab2018sagdeev}
\bibinfo{author}{Hosseini~Jenab, S.}, \bibinfo{author}{Spanier, F.} \&
  \bibinfo{author}{Brodin, G.}
\newblock \bibinfo{journal}{\bibinfo{title}{A study of the stability properties
  of {Sagdeev} solutions in the ion-acoustic regime using kinetic
  simulations}}.
\newblock {\emph{\JournalTitle{Physics of Plasmas}}}
  \textbf{\bibinfo{volume}{25}}, \bibinfo{pages}{072304}
  (\bibinfo{year}{2018}).

\bibitem{jenab2019headon}
\bibinfo{author}{Hosseini~Jenab, S.} \& \bibinfo{author}{Brodin, G.}
\newblock \bibinfo{journal}{\bibinfo{title}{Head-on collision of nonlinear
  solitary solutions to {Vlasov-Poisson} equations}}.
\newblock {\emph{\JournalTitle{Physics of Plasmas}}}
  \textbf{\bibinfo{volume}{26}}, \bibinfo{pages}{022303}
  (\bibinfo{year}{2019}).

\bibitem{turikov1984electron}
\bibinfo{author}{Turikov, V.}
\newblock \bibinfo{journal}{\bibinfo{title}{Electron phase space holes as
  localized bgk solutions}}.
\newblock {\emph{\JournalTitle{Physica Scripta}}}
  \textbf{\bibinfo{volume}{30}}, \bibinfo{pages}{73} (\bibinfo{year}{1984}).

\bibitem{bujarbarua1981theory}
\bibinfo{author}{Bujarbarua, S.} \& \bibinfo{author}{Schamel, H.}
\newblock \bibinfo{journal}{\bibinfo{title}{Theory of finite-amplitude electron
  and ion holes}}.
\newblock {\emph{\JournalTitle{Journal of Plasma Physics}}}
  \textbf{\bibinfo{volume}{25}}, \bibinfo{pages}{515--529}
  (\bibinfo{year}{1981}).

\bibitem{muschietti1999analysis}
\bibinfo{author}{Muschietti, L.}, \bibinfo{author}{Roth, I.},
  \bibinfo{author}{Ergun, R.} \& \bibinfo{author}{Carlson, C.}
\newblock \bibinfo{journal}{\bibinfo{title}{Analysis and simulation of {BGK}
  electron holes}}.
\newblock {\emph{\JournalTitle{Physics of Fluids nonlinear processes in
  geophysics}}}  (\bibinfo{year}{1999}).

\bibitem{muschietti1999phase}
\bibinfo{author}{Muschietti, L.}, \bibinfo{author}{Ergun, R.},
  \bibinfo{author}{Roth, I.} \& \bibinfo{author}{Carlson, C.}
\newblock \bibinfo{journal}{\bibinfo{title}{Phase-space electron holes along
  magnetic field lines}}.
\newblock {\emph{\JournalTitle{Geophysical research letters}}}
  \textbf{\bibinfo{volume}{26}}, \bibinfo{pages}{1093--1096}
  (\bibinfo{year}{1999}).

\bibitem{nishida1980oblique}
\bibinfo{author}{Nishida, Y.} \& \bibinfo{author}{Nagasawa, T.}
\newblock \bibinfo{journal}{\bibinfo{title}{Oblique collision of plane
  ion-acoustic solitons}}.
\newblock {\emph{\JournalTitle{Physical Review Letters}}}
  \textbf{\bibinfo{volume}{45}}, \bibinfo{pages}{1626} (\bibinfo{year}{1980}).

\bibitem{verheest2012head}
\bibinfo{author}{Verheest, F.}, \bibinfo{author}{Hellberg, M.~A.} \&
  \bibinfo{author}{Hereman, W.~A.}
\newblock \bibinfo{journal}{\bibinfo{title}{Head-on collisions of electrostatic
  solitons in nonthermal plasmas}}.
\newblock {\emph{\JournalTitle{Physical Review E}}}
  \textbf{\bibinfo{volume}{86}}, \bibinfo{pages}{036402}
  (\bibinfo{year}{2012}).

\bibitem{nakamura1999oblique}
\bibinfo{author}{Nakamura, Y.}, \bibinfo{author}{Bailung, H.} \&
  \bibinfo{author}{Lonngren, K.}
\newblock \bibinfo{journal}{\bibinfo{title}{Oblique collision of modified
  korteweg--de vries ion-acoustic solitons}}.
\newblock {\emph{\JournalTitle{Physics of Plasmas}}}
  \textbf{\bibinfo{volume}{6}}, \bibinfo{pages}{3466--3470}
  (\bibinfo{year}{1999}).

\bibitem{marchant2002asymptotic}
\bibinfo{author}{Marchant, T.~R.}
\newblock \bibinfo{journal}{\bibinfo{title}{Asymptotic solitons for a
  higher-order modified {Korteweg-de Vries} equation}}.
\newblock {\emph{\JournalTitle{Physical Review E}}}
  \textbf{\bibinfo{volume}{66}}, \bibinfo{pages}{046623}
  (\bibinfo{year}{2002}).

\bibitem{demiray2007interactions}
\bibinfo{author}{Demiray, H.}
\newblock \bibinfo{journal}{\bibinfo{title}{Interactions of nonlinear
  ion-acoustic waves in a collisionless plasma}}.
\newblock {\emph{\JournalTitle{J. Computational \& Applied Math}}}
  \textbf{\bibinfo{volume}{206}}, \bibinfo{pages}{826--831}
  (\bibinfo{year}{2007}).

\bibitem{Dubinov}
\bibinfo{author}{Dubinov, A.}, \bibinfo{author}{Kolotkov, D.~Y.} \&
  \bibinfo{author}{Sazonkin, M.}
\newblock \bibinfo{journal}{\bibinfo{title}{Supernonlinear waves in plasma}}.
\newblock {\emph{\JournalTitle{Plasma physics reports}}}
  \textbf{\bibinfo{volume}{38}}, \bibinfo{pages}{833--844}
  (\bibinfo{year}{2012}).

\bibitem{Verheest2013}
\bibinfo{author}{Verheest, F.}, \bibinfo{author}{Hellberg, M.~A.} \&
  \bibinfo{author}{Kourakis, I.}
\newblock \bibinfo{journal}{\bibinfo{title}{Electrostatic supersolitons in
  three-species plasmas}}.
\newblock {\emph{\JournalTitle{Physics of Plasmas (1994-present)}}}
  \textbf{\bibinfo{volume}{20}}, \bibinfo{pages}{012302}
  (\bibinfo{year}{2013}).

\bibitem{Saha2020}
\bibinfo{author}{Saha, A.}, \bibinfo{author}{Chatterjee, P.} \&
  \bibinfo{author}{Banerjee, S.}
\newblock \bibinfo{journal}{\bibinfo{title}{An open problem on supernonlinear
  waves in a two-component maxwellian plasma}}.
\newblock {\emph{\JournalTitle{The European Physical Journal Plus}}}
  \textbf{\bibinfo{volume}{135}}, \bibinfo{pages}{1--8} (\bibinfo{year}{2020}).

\bibitem{hakim2020continuum}
\bibinfo{author}{Hakim, A.~H.} \emph{et~al.}
\newblock \bibinfo{journal}{\bibinfo{title}{Continuum electromagnetic
  gyrokinetic simulations of turbulence in the tokamak scrape-off layer and
  laboratory devices}}.
\newblock {\emph{\JournalTitle{Physics of Plasmas}}}
  \textbf{\bibinfo{volume}{27}}, \bibinfo{pages}{042304}
  (\bibinfo{year}{2020}).

\bibitem{juno2018discontinuous}
\bibinfo{author}{Juno, J.}, \bibinfo{author}{Hakim, A.},
  \bibinfo{author}{TenBarge, J.}, \bibinfo{author}{Shi, E.} \&
  \bibinfo{author}{Dorland, W.}
\newblock \bibinfo{journal}{\bibinfo{title}{Discontinuous galerkin algorithms
  for fully kinetic plasmas}}.
\newblock {\emph{\JournalTitle{Journal of Computational Physics}}}
  \textbf{\bibinfo{volume}{353}}, \bibinfo{pages}{110--147}
  (\bibinfo{year}{2018}).

\bibitem{hakim2019conservative}
\bibinfo{author}{Hakim, A.}, \bibinfo{author}{Francisquez, M.},
  \bibinfo{author}{Juno, J.} \& \bibinfo{author}{Hammett, G.~W.}
\newblock \bibinfo{journal}{\bibinfo{title}{Conservative discontinuous galerkin
  schemes for nonlinear dougherty-fokker-planck collision operators}}.
\newblock {\emph{\JournalTitle{Journal of Plasma Physics}}}
  \textbf{\bibinfo{volume}{86}}, \bibinfo{pages}{905860403},
  \doiprefix\url{10.1017/S0022377820000586} (\bibinfo{year}{2020}).

\bibitem{hakim2020alias}
\bibinfo{author}{Hakim, A.} \& \bibinfo{author}{Juno, J.}
\newblock \bibinfo{journal}{\bibinfo{title}{Alias-free, matrix-free, and
  quadrature-free discontinuous galerkin algorithms for (plasma) kinetic
  equations}}.
\newblock {\emph{\JournalTitle{arXiv preprint arXiv:2004.09019}}}
  (\bibinfo{year}{2020}).

\bibitem{arnold2011serendipity}
\bibinfo{author}{Arnold, D.~N.} \& \bibinfo{author}{Awanou, G.}
\newblock \bibinfo{journal}{\bibinfo{title}{The serendipity family of finite
  elements}}.
\newblock {\emph{\JournalTitle{Foundations of Computational Mathematics}}}
  \textbf{\bibinfo{volume}{11}}, \bibinfo{pages}{337--344}
  (\bibinfo{year}{2011}).

\bibitem{juno2020vlasovpic}
\bibinfo{author}{Juno, J.}, \bibinfo{author}{Swisdak, M.~M.},
  \bibinfo{author}{Tenbarge, J.~M.}, \bibinfo{author}{Skoutnev, V.} \&
  \bibinfo{author}{Hakim, A.}
\newblock \bibinfo{journal}{\bibinfo{title}{Noise-induced magnetic field
  saturation in kinetic simulations}}.
\newblock {\emph{\JournalTitle{Journal of Plasma Physics}}}
  \textbf{\bibinfo{volume}{86}}, \bibinfo{pages}{175860401},
  \doiprefix\url{10.1017/S0022377820000707} (\bibinfo{year}{2020}).

\bibitem{courant1928partiellen}
\bibinfo{author}{Courant, R.}, \bibinfo{author}{Friedrichs, K.} \&
  \bibinfo{author}{Lewy, H.}
\newblock \bibinfo{journal}{\bibinfo{title}{{\"U}ber die partiellen
  differenzengleichungen der mathematischen physik}}.
\newblock {\emph{\JournalTitle{Mathematische annalen}}}
  \textbf{\bibinfo{volume}{100}}, \bibinfo{pages}{32--74}
  (\bibinfo{year}{1928}).

\bibitem{courant1967partial}
\bibinfo{author}{Courant, R.}, \bibinfo{author}{Friedrichs, K.} \&
  \bibinfo{author}{Lewy, H.}
\newblock \bibinfo{journal}{\bibinfo{title}{On the partial difference equations
  of mathematical physics}}.
\newblock {\emph{\JournalTitle{IBM journal of Research and Development}}}
  \textbf{\bibinfo{volume}{11}}, \bibinfo{pages}{215--234}
  (\bibinfo{year}{1967}).

\bibitem{vasyliunas1968survey}
\bibinfo{author}{Vasyliunas, V.~M.}
\newblock \bibinfo{journal}{\bibinfo{title}{A survey of low-energy electrons in
  the evening sector of the magnetosphere with ogo 1 and ogo 3}}.
\newblock {\emph{\JournalTitle{Journal of Geophysical Research}}}
  \textbf{\bibinfo{volume}{73}}, \bibinfo{pages}{2839--2884}
  (\bibinfo{year}{1968}).

\bibitem{pierrard2010kappa}
\bibinfo{author}{Pierrard, V.} \& \bibinfo{author}{Lazar, M.}
\newblock \bibinfo{journal}{\bibinfo{title}{Kappa distributions: theory and
  applications in space plasmas}}.
\newblock {\emph{\JournalTitle{Solar Physics}}} \textbf{\bibinfo{volume}{267}},
  \bibinfo{pages}{153--174} (\bibinfo{year}{2010}).

\bibitem{summers1991modified}
\bibinfo{author}{Summers, D.} \& \bibinfo{author}{Thorne, R.~M.}
\newblock \bibinfo{journal}{\bibinfo{title}{The modified plasma dispersion
  function}}.
\newblock {\emph{\JournalTitle{Physics of Fluids B: Plasma Physics}}}
  \textbf{\bibinfo{volume}{3}}, \bibinfo{pages}{1835--1847}
  (\bibinfo{year}{1991}).

\bibitem{hellberg2009comment}
\bibinfo{author}{Hellberg, M.}, \bibinfo{author}{Mace, R.},
  \bibinfo{author}{Baluku, T.}, \bibinfo{author}{Kourakis, I.} \&
  \bibinfo{author}{Saini, N.}
\newblock \bibinfo{journal}{\bibinfo{title}{Comment on “mathematical and
  physical aspects of kappa velocity distribution”[phys. plasmas 14, 110702
  (2007)]}}.
\newblock {\emph{\JournalTitle{Physics of Plasmas}}}
  \textbf{\bibinfo{volume}{16}}, \bibinfo{pages}{094701}
  (\bibinfo{year}{2009}).

\bibitem{cairns1995electrostatic}
\bibinfo{author}{Cairns, R.} \emph{et~al.}
\newblock \bibinfo{journal}{\bibinfo{title}{Electrostatic solitary structures
  in non-thermal plasmas}}.
\newblock {\emph{\JournalTitle{Geophysical Research Letters}}}
  \textbf{\bibinfo{volume}{22}}, \bibinfo{pages}{2709--2712}
  (\bibinfo{year}{1995}).

\bibitem{kazeminezhad2003vlasov}
\bibinfo{author}{Kazeminezhad, F.}, \bibinfo{author}{Kuhn, S.} \&
  \bibinfo{author}{Tavakoli, A.}
\newblock \bibinfo{journal}{\bibinfo{title}{Vlasov model using kinetic phase
  point trajectories}}.
\newblock {\emph{\JournalTitle{Physical Review E}}}
  \textbf{\bibinfo{volume}{67}}, \bibinfo{pages}{026704}
  (\bibinfo{year}{2003}).

\end{thebibliography}
\end{document}